# The Critical Coronal Transition Region:
# A Physics-framed Strategy to Uncover the Genesis of the Solar Wind and Solar Eruptions

A. Vourlidas[1], A. Caspi[2], Y.-K. Ko[3], J. M. Laming[3], J. Mason[1], M. P. Miralles[4], N-E Raouafi[1], J. C. Raymond[4], D. Seaton[2], L. Strachan[3], N. Viall[5], J. Vievering[1], M. West[2]

[1]*Johns Hopkins University Applied Physics Laboratory,* [2]*Southwest Research Institute, Boulder, CO,* [3]*U.S. Naval Research Laboratory, Washington, DC,* [4]*Center for Astrophysics | Harvard & Smithsonian, Cambridge, MA,* [5]*NASA Goddard Space Flight Center, Greenbelt, MD*

## Synopsis

Our current theoretical and observational understanding suggests that critical properties of the solar wind and Coronal Mass Ejections (CMEs) are imparted within 10 Rs, particularly below 4 Rs. This seemingly narrow spatial region encompasses the transition of coronal plasma processes through the entire range of physical regimes from fluid to kinetic, and from primarily closed to open magnetic field structures. From a physics perspective, therefore, it is more appropriate to refer to this region as the **Critical Coronal Transition Region (CCTR)** to emphasize its physical, rather than spatial, importance to key Heliophysics science.

This white paper argues that the comprehensive exploration of the CCTR will answer two of the most central Heliophysics questions, '*How and where does the solar wind form?*' and *'How do eruptions form?',* by unifying hardware/software/modeling development and seemingly disparate research communities and frameworks. We describe the outlines of decadal-scale plan to achieve that by 2050.

# The Current State

Long the source of fascination for humanity, the solar corona[1] has increasingly become a central target of Heliophysics research as space-based research became to dominate the field. The corona is the source of both the *ambient* and *eruptive* solar wind (SW). Thus, the effects of physical processes that operate in the corona permeate every aspect of Heliophysics, from solar to space physics to the study of planetary magnetospheres, ionospheres, and atmospheres to the interaction of our heliosphere with interstellar space.

It is no surprise, then, that much of the current research effort and hardware investment in Heliophysics revolves around variations of a single (two-sided) overarching question: *How does the solar wind (ambient and eruptive) form?* The question has remained open ever since the discoveries of the solar wind in 1962 and of Coronal Mass Ejections (CMEs) in 1971.

We have made progress in the last half-century but despite the energy and funds expended on these problems, the fundamental questions on how the solar wind and CMEs form remain open. **Table 1** attempts to summarize, at a high level, the current status (Viall et al. 2023a). Evidently, both are complex problems but we know enough to envision a complete solution within the next 30 years. This White Paper (WP) describes the broad outlines of a 30-yr plan to resolve these issues. Linked WPs (e.g., Seaton et al. 2023) with more detailed descriptions of some key science investigations are referenced where necessary.

| Table 1 Top-level Synopsis of the State of Knowledge of two Major Questions in Heliophysics as of 2022. | |
|---|---|
| **How and where does the solar wind form?** | |
| **Current Status (top-level)** | **Issues** |
| - Large-scale magnetic field is responsible for the structure of the corona<br>- General picture is that the SW is bimodal SW: Fast SW (FSW) from open field coronal holes. Slow SW (SSW) from everywhere else, including closed fields.<br>- There are a plethora of observations of corona and solar wind that do not fit the bimodal SW picture<br>- SW energization can come from waves, turbulence, reconnection, instabilities, or combinations of multiple processes. | - No routine measurements of the physical properties (temperature, magnetic field, speed, composition, waves, turbulence, etc.) in the critical range ~1.2–5 Rs<br>- Reliance on incomplete measurements of the photospheric field and ad-hoc prescriptions of coronal heating to model the corona<br>- Small-scale structure and variability in the low atmosphere have wide impacts in the corona (e.g., S-web, supergranules)<br>- Complex magnetic expansion and connectivity from the photosphere through the source of the solar wind<br>- Bimodal SW picture does not capture time-dynamics, 3D complexity, or multi-scale feedback that are critical to the formation of the solar wind |
| **How do eruptions form?** | |
| **Current Status (top-level)** | **Issues** |
| - CMEs (and flares) are coronal phenomena powered by the release of magnetic energy via reconnection<br>- CMEs are ejections of magnetic flux ropes and develop within 1–15 Rs<br>- CME acceleration/speed peak mostly <3 Rs / 20 Rs, resp.<br>- Similar processes drive eruptions over a wide spatio-temporal range, from jets (arcsecs/arcmins) to streamer-blowouts (>90º/days)<br>- High-energy SEPs accelerated below 10 Rs | - Impossible to predict eruption due to incomplete understanding of energy accumulation, role of instabilities, triggers<br>- CME magnetic content unknown due to: lack of $B_{cor}$ measurements; incomplete understanding of CME inner corona evolution and force balance; no thermal information<br>- Incomplete understanding of CME early evolution (rotation, deflection, compression)<br>- No routine measurements of physical properties of shocks and plasma prevent closure on particle acceleration processes |

---

[1] We define the corona as the region between two physical interfaces of the solar atmosphere: from $\beta>1$ to $\beta<1$ interface (~upper chromosphere) to the sub- to super-Alfvènic solar wind interface (~20 Rs, nominally).

## Research 'Choke Points' and Path Forward

The issues in Table 1 readily identify several 'research chokepoints' that hinder progress (see Seaton et al. 2023 for additional details):

- **Incomplete measurements.** This is the key chokepoint. The physical properties of the corona above ~1.2 Rs are rarely measured, particularly temperature, magnetic field, flow speed, and composition. The various regions are covered by different instruments in different wavelengths (low corona in the EUV, middle corona in the visible). The photospheric magnetic field is usable only within 60º from central meridian.
- **Multi-scale processes and feedback**: coronal physics encompasses a large range of temporal and spatial scales crossing several physical interfaces (see Kepko et al. 2023a for details). The system is interconnected with feedback from small to large scales and vice versa, including long-range interactions.
- **Fragmented community:** The corona is observed across a range of wavelengths (radio to X-rays) and with different instrumentation (both remote and in-situ) not always operating concurrently. This has fragmented the community into distinct plasma regimes, largely clustering around regions where existing instrumentation has made observations widely available and where models can be sufficiently self-contained to be tractable. This disunity precipitates stovepiping of research topics, stifling cross-disciplinary innovation required to understand the system as a whole.

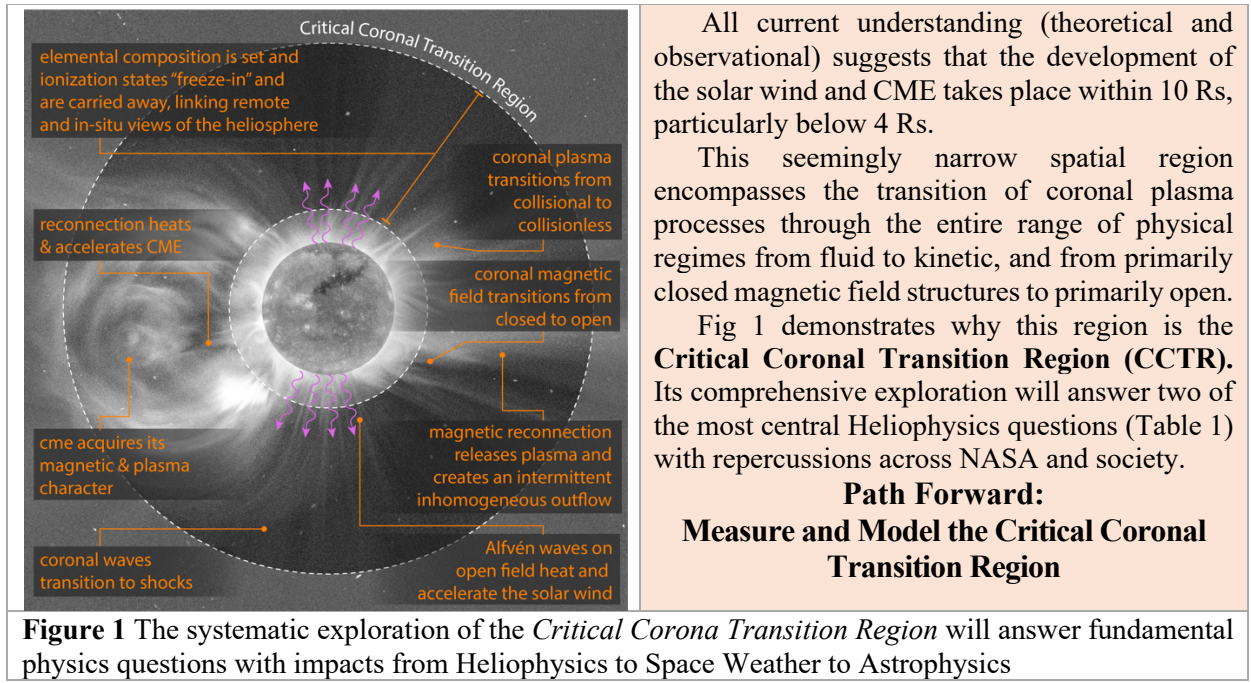

All current understanding (theoretical and observational) suggests that the development of the solar wind and CME takes place within 10 Rs, particularly below 4 Rs.

This seemingly narrow spatial region encompasses the transition of coronal plasma processes through the entire range of physical regimes from fluid to kinetic, and from primarily closed magnetic field structures to primarily open.

Fig 1 demonstrates why this region is the **Critical Coronal Transition Region (CCTR).** Its comprehensive exploration will answer two of the most central Heliophysics questions (Table 1) with repercussions across NASA and society.

**Path Forward:
Measure and Model the Critical Coronal Transition Region**

**Figure 1** The systematic exploration of the *Critical Corona Transition Region* will answer fundamental physics questions with impacts from Heliophysics to Space Weather to Astrophysics

## Achieving Closure by 2050

To accelerate scientific discovery, we must adopt a systems approach with coordinated development along three axes:

- **Better tools** (data assimilation, big data techniques, model dissemination, virtual repositories)
- **Better measurements** (the right measurements from the right location for the right duration; Vourlidas et al. 2023)
- **United community** (unify the research language, expand premise of DRIVE Science Centers, enhance interdisciplinary & international collaborations [Kepko et al. 2023b], train system-minded researchers).

These developments feed and strengthen the two pillars of the research enterprise–-observations and theory/modeling---through a chain of interlinked science investigations as outlined in our indicative 30-year plan in **Table 2.**

**Table 2** Indicative 30-year plan for Achieving Closure on the Outstanding Scientific Problems of Solar Wind and CME Formation.
The black box outlines the scope for this Decadal Survey although several of the goals and investigations in the next decade could be brought forward depending on research and funding priorities.

| Goals | Questions | 0–10 years | | 10–20 years | | 20–30 years |
|---|---|---|---|---|---|---|
| | | Investigation | Goals | Investigations | Goals | Goals |
| How and where does the Solar Wind Form?[2] | - Where does the SW originate?<br>- How is the SW released and accelerated?<br>- What determines the SW composition and ionic state?<br>- What is the origin & evolution of the mesoscale plasma & magnetic field structure?<br>- What is the origin of Alfvénic fluctuations?<br>- How is SW turbulence driven and dissipated?<br>- How do the SW kinetic distribution functions evolve? | - Trace magnetic & plasma connectivity from photosphere through the corona<br>- Measure physical properties of CCTR[3]<br>- Weed-out theories of SW heating/acceleration (test/validate predictions)<br>- Advance coronal modeling<br>- Investigate the wave/turbulence properties of coronal plasmas<br>- Advance coronal magnetometry methods | - Connect abundance variation in the Sun and solar wind<br>- Link small-scale structures from chromosphere through corona to heliosphere<br>- Improve physical description of the corona<br>- Map the wave properties of coronal plasma<br>- Data-driven time-dependent coronal modeling | - Trace energy & plasma flow throughout the corona<br>- Trace magnetic connectivity throughout the corona<br>- Measure the plasma and magnetic field of the closed-field solar atmosphere with in situ probes[4] | - Regular polar coverage of B-field & corona[5]<br>- First sustained multi-height B-field through the $\beta=1$ layer.<br>- $2\pi+$ corona/B-field coverage | - Achieve full coverage of the Sun in a few key regimes<br>- Closure on the two overarching goals<br>- Move towards prediction of solar eruptive activity<br>- Move towards integration of solar knowledge to stellar systems |
| | Tools | - Ingest/analyze multi-platform/view data<br>- Ingest multi-view B-field data<br>- Deploy Big Data methods<br>- Begin 'fusing' remote/in-situ methodologies<br>- Facilitate access/use of advanced models by the community | | - Booms/formation-flying; EUV/UV coatings; imaging spectropolarimetry<br>- High resolution X-ray imaging<br>- ML/AI use to long-term solar dbase fusion<br>- Practical multi-scale kinetic/MHD simulations<br>- time-dependent data-driven simulations<br>- Fusion of remote/in-situ methodologies | | Routine data assimilation |
| | Measurements | - Physical properties of coronal and CME plasma with multiple instruments (e.g., PSP-SO-DKIST)<br>- Measure coronal magnetic field<br>- Off-limb spectroscopy, high-res imaging (VIS/EUV),<br>- First polar, off Sun Earth line magnetic field | | - $2\pi$ $B_{phot}$ coverage<br>- High-sensitivity spectroscopy for 'seed' detection<br>- Multi-height B measurements<br>- Time/space-resolved 3D imaging (polarization or tomography) | | $4\pi$ $B_{phot}$ & corona coverage |
| | Community | - Expand remit of Drive Science Centers; FSTs, ISTPNext | | - Establish 'Corona-as-a-system' research program | | |
| How do CMEs form? | - What are the roles of ideal/non-ideal processes in CME eruption?<br>- Where does the CME magnetic field originate?<br>- What is the energy budget and distribution in eruptions?<br>- How do coronal shocks form and accelerate particles? | - particle acceleration & shock development <10 Rs<br>- 3D CME evolution <10 Rs<br>- Eruption Energy budgets<br>- Estimate CME magnetic content<br>- Feedback to surface (X-rays, γ-rays, CME effects on global B-field) | - Reliable Identification of SEP origins in individual events<br>- Remote measurements of 'seed' populations<br>- Improve coronal shock modeling<br>- Advance description of forces acting on CMEs<br>- quantitative $B_{CME}$ estimate0<br>- SEP production at CME shocks | - particle acceleration & shock tracking <10 Rs<br>- 3D CME evolution <10 Rs<br>- detailed eruption energy budgets<br>- Trace $B_{COR}$ CME transformation and subsequent evolution | - Measure coronal currents in ARs<br>- Semi-operational coronal MHD modeling<br>- Operational SEP prediction | - Deploy CME prediction methods<br>- Reliable mapping of $B_{cor}$ to Alfvénic zone<br>- Operational coronal modeling |

---

[2] See Viall et al. (2023)
[3] See Rabin et al. (2023)
[4] See Viall et al., (2023b)
[5] For example, Raouafi et al. (2023)